\documentclass[iop,apj,numberedappendix]{emulateapj}
\usepackage[latin1]{inputenc}
\usepackage{amssymb}
\usepackage{amsmath}
\usepackage{graphicx}
\usepackage{xspace}
\usepackage{natbib}
\usepackage{url}
\usepackage{paralist}
 \usepackage{aasmacros}
\usepackage{multirow}

\slugcomment{ }

\shorttitle{X-ray reflection in 4C 74.26}
\shortauthors{Lohfink et al.}

\begin{document}

\title{The X-ray reflection spectrum of the radio-loud quasar 4C\,74.26}


\author{Anne M. Lohfink\altaffilmark{1}, Andrew C. Fabian\altaffilmark{1}, David R. Ballantyne\altaffilmark{2}, S.E. Boggs\altaffilmark{3}, Peter Boorman\altaffilmark{4}, F.E. Christensen\altaffilmark{5}, W.W. Craig\altaffilmark{5,6}, Duncan Farrah\altaffilmark{7}, Javier Garc\'{i}a\altaffilmark{8,9,10}, C.J. Hailey\altaffilmark{11}, F.A. Harrison\altaffilmark{8}, Claudio Ricci\altaffilmark{12}, Daniel Stern\altaffilmark{13}, W.W. Zhang\altaffilmark{14}}
\email{alohfink@ast.cam.ac.uk}
\altaffiltext{1}{Institute of Astronomy, University of Cambridge, Madingley Road, Cambridge CB3 0HA, UK}
\altaffiltext{2}{Center for Relativistic Astrophysics, School of Physics, Georgia Institute of Technology, 837 State Street, Atlanta, GA 30332-0430, USA}
\altaffiltext{3}{Space Sciences Laboratory, University of California, Berkeley, CA 94720-7450, USA}
\altaffiltext{4}{Department of Physics \& Astronomy, Faculty of Physical Sciences and Engineering, University of Southampton, Southampton, SO17 1BJ,UK}
\altaffiltext{5}{DTU Space-National Space Institute, Technical University of Denmark, Elektrovej 327, DK-2800 Lyngby, Denmark}
\altaffiltext{6}{Lawrence Livermore National Laboratory, Livermore, CA 94550, USA}
\altaffiltext{7}{Department of Physics, Virginia Tech, Blacksburg, VA 24061, USA}
\altaffiltext{8}{Cahill Center for Astronomy and Astrophysics, Caltech, Pasadena, CA 91125, USA}
\altaffiltext{9}{Dr.-Karl-Remeis-Sternwarte \& ECAP, Universit{\"a}t Erlangen-N{\"u}rnberg, Sternwartstrasse 7, 96049, Bamberg, Germany}
\altaffiltext{10}{Harvard-Smithsonian Center for Astrophysics, 60 Garden St., Cambridge, MA 02138, USA}
\altaffiltext{11}{Columbia Astrophysics Laboratory, Columbia University, New York, NY 10027, USA}
\altaffiltext{12}{Instituto de Astrof\'{i}sica and Centro de Astroingenier\'{i}a, Facultad de F\'{i}ica, Pontificia Universidad Cat\'{o}lica de Chile, Casilla 306, Santiago 22, Chile}
\altaffiltext{13}{Jet Propulsion Laboratory, California Institute of Technology, Pasadena, CA 91109, USA}
\altaffiltext{14}{X-ray Astrophysics Laboratory, NASA Goddard Space Flight Center, Greenbelt, MD 20771, USA}

\begin{abstract}
The relativistic jets created by some active galactic nuclei are important agents of AGN feedback. In spite of this, our understanding of what produces these jets is still incomplete. X-ray observations, which can probe the processes operating in the central regions in immediate vicinity of the supermassive black hole, the presumed jet launching point, are potentially particularly valuable in illuminating the jet formation process. Here, we present the hard X-ray \textit{NuSTAR} observations of the radio-loud quasar 4C\,74.26 in a joint analysis with quasi-simultaneous, soft X-ray \textit{Swift} observations. Our spectral analysis reveals a high-energy cut-off of 183$_{-35}^{+51}$\,keV and confirms the presence of ionized reflection in the source. From the average spectrum we detect that the accretion disk is mildly recessed with an inner radius of $R_\mathrm{in}=4-180\,R_\mathrm{g}$. However, no significant evolution of the inner radius is seen during the three months covered by our \textit{NuSTAR} campaign. This lack of variation could mean that the jet formation in this radio-loud quasar differs from what is observed in broad-line radio galaxies. 
\end{abstract}



\section{Introduction}\label{intro}
Radio-loud active galactic nuclei (AGN) are interesting objects to study as their relativistic jets deposit energy into the interstellar medium. This is known to have a profound influence on star formation and the development of the galaxy \citep[e.g.,][]{Fabian2012,Wagner2015}. Despite this importance it is still not clear why only some AGN display jets. A study focus has been broad-line radio galaxies, the radio-loud counterparts of broad line Seyfert~1s. Previous observations of broad line radio galaxies have revealed that the innermost regions are where the processes of jet formation manifest themselves \citep{Marscher2002,Chatterjee2009,Chatterjee2011,Lohfink2013}. It is thought that the ejection of a new jet knot into the jet in those sources is preceded by a disruption of the inner parts of the accretion disk most likely due to changes to the magnetic fields threading the accretion disk \citep{Sikora2013b}. This disruption can be observed by studying the coronal emission that illuminates the accretion disk and its reprocessed emission, the so-called reflection spectrum. If the accretion disk is disrupted, no or little backscattered emission is detected from the disrupted region and the accretion disk appears truncated. However, such an observation of a truncated disk could only be directly connected with the jet formation process on one occasion \citep{Lohfink2013}. This might be mainly because of the need for simultaneous radio and X-ray observations to follow the changes in the jet, given that ionized reflection from the accretion disk has been detected in several other radio-loud sources, for example, in 3C\,390.3 \citep{Lohfink2015} and the subject of this paper 4C\,74.26 \citep{Ballantyne2005,Ballantyne2005a,Larsson2008}. 

With a bolometric luminosity of $2\times 10^{46}\,\mathrm{ergs}\,\mathrm{s}^{-1}$ \citep{Woo2002}, 4C\,74.26 is one of the nearest powerful radio-loud quasars ($z=0.104$) and its broad Fe K$\alpha$ line was first discovered in a 2004 \textit{XMM-Newton} observation \citep{Ballantyne2005,Ballantyne2005a}. This was later confirmed from a \textit{Suzaku} observation in 2007, in which the data suggested that the accretion disk was possibly recessed, i.e. does not extend all the way to its innermost stable circular orbit \citep{Larsson2008}. Besides the ionized reflection component and its associated broad iron line, the X-ray spectrum of 4C\,74.26 has been found to be typical of a broad line radio galaxy consisting of a relatively flat power law continuum with weak neutral reflection. Moreover, the power law continuum in 4C\,74.26 has an observed high-energy cut-off. \textit{BeppoSAX} constrained the cut-off to $145^{+294}_{-66}$\,keV \citep{Grandi2005} and a joint \textit{INTEGRAL}-ISGRI/\textit{Swift}-BAT analysis further improved the constraint to $190^{+18}_{-66}$\,keV \citep{Malizia2014a}. The spectral emission components are attenuated by Galactic absorption, excess neutral absorption above the Galactic value, and a warm absorber \citep{Ballantyne2005,DiGesu2016}. \citet{Tombesi2014} also detected signs of a highly ionized ultra-fast outflow in the \textit{XMM-Newton} data of the source. A similar discovery was made by \citet{Gofford2013} in \textit{Suzaku} data. 

Being radio-loud, the jets of 4C\,74.26 are also well studied, especially in the radio band. One of the key findings of this work is that the jet angle is estimated to be less than 49 degrees to the line-of-sight \citep{Pearson1992}, which assuming Bardeen-Petterson alignment of the disk and the jet can be used to check the reflection modeling results. \textit{Chandra} and \textit{XMM-Newton} imaging has also shown that the central region of the AGN has a complex X-ray structure, showing both the nucleus as well as a weak hot spot resolved in the X-ray band about 5\,arcmin ($\sim$600\,kpc) from the central source \citep{Erlund2007}. The black hole mass of  4C\,74.26 has been estimated to be $\log(M_\text{BH}/M_\odot)=9.6\pm0.5$ \citep{Woo2002}.

In this paper we study the broad-band spectrum of 4C\,74.26 with particular focus on the reflection spectrum. We begin with an explanation of how the spectra were produced in Section~\ref{datared}, which is followed by the spectral analysis of the average spectrum in Section\,\ref{spec_an} and of the spectra of the individual observations in Section\,\ref{spectral_indi}.

\section{Data Reduction}\label{datared}

\begin{figure}
\includegraphics[width=\columnwidth]{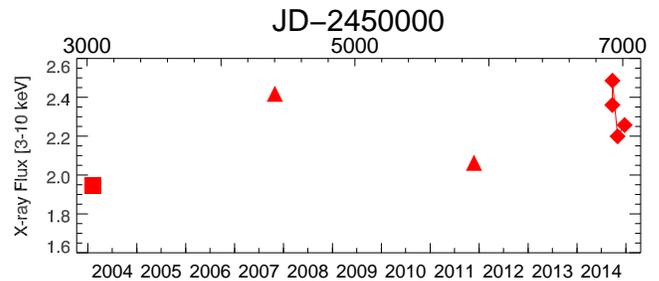}
\vspace{-8\baselineskip}
\caption{Overview light curve of the obervation-averaged 3-10\,keV X-ray flux in units of $10^{-11}\,\text{erg}\,\text{s}^{-1}\,\text{cm}^{-2}$. Squares correspond to \textit{XMM-Newton}, triangles to \textit{Suzaku} and diamonds to \textit{NuSTAR} observations.}\label{long_lightcurve}
\end{figure}

In this paper we use data obtained from the \textit{Swift} \citep{Gehrels2004}, and \textit{NuSTAR} \citep{Harrison2013} satellites. An overview of the data considered in this work is given in Table~\ref{obs}, it consists of four observations taken within three months of each other in 2014. In Figure~\ref{long_lightcurve} we show the observed 3-10\,keV fluxes of 4C\,74.26 during the observations considered in this work. The flux values of the archival \textit{Suzaku} \citep{Mitsuda2007}, and \textit{XMM-Newton} \citep{Jansen2001} observations are also included. As expected for a high-mass black hole such as 4C\,74.26, the flux changes are small with the \textit{XMM-Newton} observation displaying the lowest flux.

\begin{table}
\caption{\textit{NuSTAR} observations used in the spectral analysis}\label{obs}
\begin{tabular}{c|c|c|c}
Observatory & ObsID & Exposure [ks] & Start Date \\
\hline 
\textit{NuSTAR} & 60001080002 & 17 & 2014-09-21\\
\textit{NuSTAR} & 60001080004 & 53 & 2014-09-22\\
\textit{NuSTAR} & 60001080006 & 82 & 2014-10-30\\
\textit{NuSTAR} & 60001080008 & 39 & 2014-12-22\\
\textit{Swift} & 00080795001& 2 & 2014-09-21\\
\textit{Swift} & 00080795002 & 2 & 2014-09-24\\
\textit{Swift} & 00080795003 & 2 & 2014-10-31\\
\textit{Swift} & 00080795004 & 2 & 2014-12-22\\
\end{tabular}
\end{table}

\subsection{Swift}

All \textit{Swift} spectra have been produced using the \textit{Swift}-XRT data products generator, which follows procedures outlined in \citet{Evans2009}. In the analysis we use both the spectra from the individual observations as well as the average spectrum.

\subsection{NuSTAR}

To reduce the \textit{NuSTAR} data, the event files are first screened using \texttt{nupipeline}. To obtain better signal-to-noise at high energies we applied a stricter screening around the SAA passages than usual by flagging both time intervals with high shield single rates and unusually increased CZT detector event count rates, thereby excluding times of increased background. The spectra were then extracted using the \texttt{nuproducts} tool. The regions from which the source and background spectrum were extracted were both circular with 60 arcsec radii. 

During the fitting, the two focal plane modules, FPMA and FPMB, were fitted separately but with the same model except for a cross calibration constant. Both spectra were binned to a signal to noise ratio of 10. 

\section{Spectral Analysis}\label{spec_an}
\subsection{Exploration -- stacked NuSTAR data only}\label{section_nuonly}
We begin our analysis with an exploration of the average \textit{NuSTAR} spectrum as a guide for the later analysis including the soft X-ray data. 
\begin{figure}[ht]
\includegraphics[width=\columnwidth]{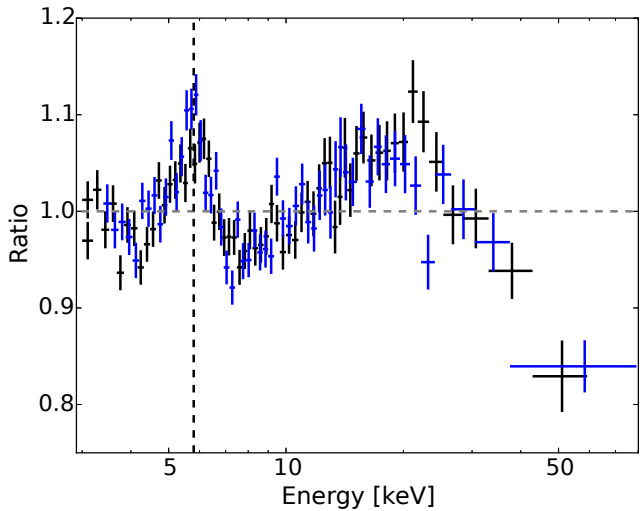}
\caption{Ratio residuals for the average \textit{NuSTAR} (FPMA [black], FPMB [blue]) spectra fitted with a simple power law continuum. The dashed vertical line indicates 6.4\,keV rest frame for comparison. The data have been rebinned for clarity.}\label{residuals_pownuonly}
\end{figure}
At first we examine the residuals to a simple power law (Fig. \ref{residuals_pownuonly}). The ratio residuals show clear indications for the presence of reflection in the spectrum. Both the Compton hump and the iron line, centered at approximately 6.4\,keV rest frame, are prominent features in the spectrum. Besides reflection, no other strong features are apparent. Based on this observation, we first attempt to describe our average \textit{NuSTAR} spectra with a combination of a power law continuum and a cold reflector attenuated by Galactic absorption. The \texttt{pexmon} model is able to describe both the power law continuum as well as the distant reflection. The strength of the reflection with respect to the power law is controlled by the reflection fraction parameter ($R$). The inclination and iron abundance ($A_\text{Fe}$ in solar units) of the reflector are also parameters of the model. For our fits, we limit the inclination to less than 49\,degrees as it was determined in the previous radio studies \citep{Pearson1992}, assuming the cold reflection stems from the outermost parts of the accretion disk. Finally, we keep the high energy cut-off of the power law continuum fixed to 1000\,keV initially. We model the Galactic absorption with the \texttt{TBnew} model \footnote{http://pulsar.sternwarte.uni-erlangen.de/wilms/research/tbabs/}, with abundances set to those of \citet{Wilms2000}. During the fitting the Galactic absorption column is kept fixed\footnote{https://heasarc.gsfc.nasa.gov/cgi-bin/Tools/w3nh/w3nh.pl} at a value of $1.15\times10^{20}\,\text{cm}^{-2}$.  

\begin{figure}[ht]
\includegraphics[width=\columnwidth]{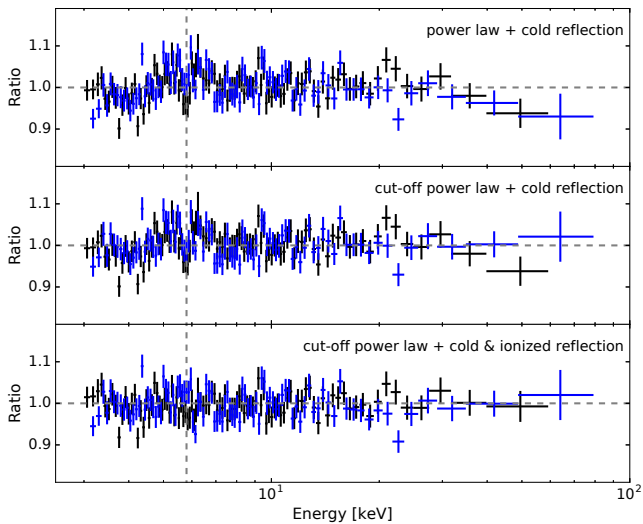}
\caption{Ratio residuals for the average \textit{NuSTAR} (FPMA [black], FPMB [blue]) spectra fitted with a power law continuum and cold reflector (Top panel), a cut-off power law continuum and cold reflector (Middle panel), and a cut-off powerlaw with cold and blurred ionized reflector (Bottom panel). The vertical, dashed, grey line indicates the position of the neutral Fe K line in the source. The residuals have been rebinned for clarity.}\label{residuals_nustaronly}
\end{figure}

While the fit is able to describe the data roughly ($\chi^2=977.4$ for 838 dof), clear residuals remain (Fig.~\ref{residuals_nustaronly}, top panel). Including a high-energy turnover leads to an improvement of $\Delta\chi^2=40.6$ for -1 dof (Fig.~\ref{residuals_nustaronly}, middle panel). Even when the high energy cut-off is included residuals remain in the iron line region, indicating the iron line could be broadened. This is also supported by the measured line equivalent width of $\sim 200$\,eV\footnote{To obtain the equivalent width of the line, it was modeled with a simple Gaussian and the continuum with the model \texttt{pexrav}.} and a line width of $\sigma=0.57_{-0.08}^{+0.09}$\,keV. We therefore test whether ionized relativistic reflection can lead to a further improvement. The \texttt{relxilllp} model \citep{Garcia2014}, which describes the reflection spectrum resulting from a X-ray point source at height $h$ above the accretion disk, is used for this purpose. It includes the relativistic blurring that a reflection spectrum originating from the vicinity of a black hole would be subject to. To get the best constraints on the reflection parameters for the given model, we set the model so it calculates both the emissivity profile as well as the strength of the reflection features self-consistently for a given source height. This means the primary emission is included in this model and we fix the reflection fraction of the neutral reflection to -1 (to only contribute the reflected part to our model). The normalization of the \texttt{pexmon} model then controls the strength of the neutral reflection. Furthermore we assume that the irradiating continuum is the primary continuum for both reflection components and consequently tie the photon indexes and high energy-cut-offs between the \texttt{pexmon} and \texttt{relxilllp} models. The inclination is also assumed to be the same for both reflectors. As the \texttt{relxill} version 4.0b used here does not properly model the cosmological redshift we apply this correction separately, using the xspec model \texttt{zashift}. As \texttt{zashift} not only shifts the energies but also adjusts the flux level of the model it is applied to, we have applied it to both \texttt{relxilllp} and \texttt{pexmon} to preserve the relative normalizations. A fit including ionized reflection leads to a final $\chi^2$ of 899.8 for 833 dof and a good description of the data. The best fit parameters are shown in Table~\ref{tab_nustaronly} and the residuals in Fig.~\ref{residuals_nustaronly}, bottom panel. With a photon index of 1.84$_{-0.02}^{+0.03}$ and a rather poorly constrained source height of $25_{-19}^{+29}\,R_g$, we recover typical parameters for radio-loud AGN. We are also able to detect a high-energy cut-off at $234_{-51}^{+87}$\,keV. It is clear that the \textit{NuSTAR} spectrum by itself does not have the spectral resolution to distinguish cold and ionized reflection very well for this source, with the cold reflection now being consistent with zero. Finally we note that a search for ultrafast outflows in the average \textit{NuSTAR} spectra does not yield a detection.

To assess whether the inclusion of the blurred ionized reflection component is really justified or whether the improvement in $\chi^2$ caused by random statistical fluctuations rather than a truly more complex spectrum, we use a similar technique to that outlined in \citet{Zoghbi2015}. We simulated 1000 spectra using \texttt{xspec}'s \texttt{fakeit} command drawing randomly from the best fit neutral reflection plus continuum model parameters accounting for its $1\,\sigma$ uncertainties via the covariance matrix. These simulated spectra are then fitted using both a neutral reflection plus continuum model and a neutral \& ionized reflection plus continuum model and a $\Delta \chi^2$ is calculated for each of them. We find that none of the simulated spectra have a $\Delta \chi^2$ of 37 or larger as it is observed in our real spectrum, letting us conclude the inclusion of the additional component is most likely justified.

\begin{table*}[th]
\caption{Best fit parameters of \textit{NuSTAR}-only fits: the model consists of a cut-off power law continuum with cold and ionized reflection.}\label{tab_nustaronly}
\begin{center}
\begin{tabular}{c|c|c|c|c|c|c|c|c|c}
 $\Gamma$ & $A_\text{Fe}$ & $E_\text{fold}$ & $R_\text{ion}$ & $N_\text{pex} $ & $a$ & $i$\, & $h$ & $\log(\xi)$ & $N_\text{rel}$ \\
\hline & &[keV] & & [$10^{-3}$] & & [$^\circ$] & [$R_g$] & [erg\,cm\,s$^{-1}]$ & [$10^{-4}$] \\   
\hline \hline 1.84$_{-0.02}^{+0.03}$ &  $0.8_{-0.2}^{+0.1}$ & $234_{-51}^{+87}$ & $0.47$ & $<1.8$ & uncont. & $35_{-5}^{+11}$ & $25_{-19}^{+29}$ &  $2.7\pm 0.1$ & $2.03_{-0.15}$ \\
\end{tabular}
\end{center}
\end{table*}

\subsection{Spectral modeling including the \textit{Swift} soft X-ray data}\label{spectral_all}
So far our analysis has been focused on the hard X-ray part of the spectrum. In this section we now include the available soft X-ray data. Simultaneous \textit{Swift}-XRT snapshots available with the \textit{NuSTAR} observations are averaged to an average \textit{Swift} spectrum that is fitted jointly with the \textit{NuSTAR} data. 

The \textit{Swift} data are well described with the model used to describe \textit{NuSTAR} by itself, although we see evidence of absorption at low energies. We therefore base our spectral modeling on the lessons learned in the last section, adopting a model consisting of a cut-off power law with cold and ionized reflection. The inclusion of the soft X-ray data leads to two additions, excess neutral absorption and a warm absorber. Both absorption components have been observed previously in the source \citep[e.g.,][]{DiGesu2016} and are required to obtain a good fit to the soft X-ray data ($\chi^2=972.3$ for 898 dof). 

Our results are similar to what was found in Section~\ref{section_nuonly} for the \textit{NuSTAR}-only fits. Finally, we also test whether the accretion disk could be magnetically disrupted at times, as expected as part of the jet cycle which is believed to operate in radio-loud AGN  \citep{Marscher2002,Chatterjee2009, Lohfink2013}. To test this, we allow the inner radius of the accretion disk to vary. Allowing a free inner radius leads to an significantly improved fit ($\Delta\chi^2=5.1$ for -1 dof). The inner radius can be constrained to $3-20\,R_\text{ISCO}$, which corresponds to $4-180\,R_\text{g}$ given that the black hole spin is unconstrained between -0.998 and 0.998. The other resulting parameters are shown in Table~\ref{full_band}. 

\begin{table*}[ht]
\caption{Spectral parameters of a joint fit of all soft and hard X-ray observations of 4C\,74.26. Centered values indicate that the parameter was determined jointly from all observations. The assumed model is a combination of a cut-off power law or Comptonization continuum, cold and ionized reflection. All emission is attenuated by Galactic and intrinisic neutral absorption and a warm absorber. More details about the model set-up can be found in the text.}\label{full_band}
\begin{center}
\begin{tabular}{c|c|c|c}
& Parameter & cutoffpl continuum & nthcomp continuum\\
\hline \hline Absorption & $N_\text{H, intr.}$ [$10^{21}$\,cm$^{-2}$] & $<2.3$ & $<2.2$\\
 & $N_\text{H, wa}$ [$10^{21}$\,cm$^{-2}$] & 4.0$\pm 2.0$& 4.2$_{-2.0}^{+2.1}$  \\
 & $\log{\xi}_\text{wa}$ [erg\,cm\,s$^{-1}$] & 1.1$\pm 0.2$& 1.1$_{-0.2}^{+0.1}$ \\
\hline Continuum \&  & $\Gamma$ & 1.81$\pm 0.03$ & 1.85$_{-0.01}^{+0.03}$\\
 & $E_\text{cut}$ [keV]& 183$_{-35}^{+51}$ & --\\
&kT$_e$ [keV]& --& 46$_{-11}^{+25}$\\
Cold reflection & $N_\text{pex}$ [$10^{-3}$] & $<2.0$& 1.2$_{-1.0}^{+1.2}$ \\
\hline Ionized reflection & $h$ [$R_\text{g}$] & 7$_{-2}^{+6}$ & 6$_{-1}^{+6}$ \\
& $R_\text{in} [R_\text{ISCO}]$ & 14$_{-11}^{+6}$ & $6_{-4}^{+15}$  \\
& $a$ & uncont. & uncont.\\
& $i$ [$^\circ$] & 35$_{-9}^{+11}$ & 37$_{-10}^{+11}$\\
& $A_\text{Fe}$ & 1.0$_{-0.2}^{+0.5}$ & 1.2$_{-0.4}^{+0.7}$\\
& $\log(\xi)$ [erg\,cm\,s$^{-1}$] & 2.7$_{-0.2}^{+0.1}$ & 2.7$_{-0.2}^{+0.1}$\\
& $R_\mathrm{ion}$ & 0.46 & 0.43  \\
& $N_\text{rel}$ [$10^{-4}$] & 3.6$_{-1.4}^{+1.6}$ & 3.7$_{-0.9}^{+1.4}$ \\
\hline Cross calibration & $c_\text{FB}$ & 1.02$\pm 0.01$& 1.02$\pm 0.01$\\
& $c_\text{XRT}$ & 0.85$_{-0.04}^{+0.03}$ & 0.84$\pm 0.03$\\
\hline \hline & $\chi^2$/dof & 967.2/897  & 972.6/897\\
\end{tabular}
\end{center}
\end{table*}

\begin{figure}[ht]
\includegraphics[width=\columnwidth]{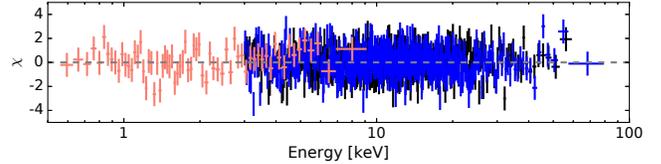}
\caption{Residuals from a fir of the average \textit{NuSTAR} (FPMA [black], FPMB [blue]) and \textit{Swift} (XRT [red]) spectra with a model, which is a combination of a Comptonization continuum, and cold and ionized reflection; all attenuated by absorption. More details about the model set-up can be found in the text.}\label{residuals_nustar_swift}
\end{figure}

With a clear detection of a high-energy cut-off, we now explore the primary X-ray continuum further. To acquire more physical information, we replace the cut-off power law with a thermal Comptonization component, modeled by \texttt{nthcomp} \citep{Zdziarski1996,Zycki1999}. In order to obtain the most self-consistent description of the data possible, we therefore replace the \texttt{relxilllp} model with the new model \texttt{relxilllpCp}. In \texttt{relxilllpCp}, the cut-off power law is replaced by \texttt{nthcomp} both for the continuum included in the model as well as that assumed for the calculation of the reflection. Instead of being described by the photon index and the exponential cut-off, the continuum is now described by a photon index, the electron temperature and the seed photon temperature (kept fixed at 0.01\,keV in our modeling). For the neutral reflection we retain the \texttt{pexmon} model with the high-energy cut-off fixed to $2\,kT_\text{e}$. While the reflection parameters remain unchanged (Table~\ref{full_band}), the measured photon index is slightly higher than for the cut-off powerlaw continuum model. The fit quality is still very good, see Fig.~\ref{residuals_nustar_swift}, although slightly worse than for the fit with cut-off power law continuum. 

\section{Time-resolved \textit{Swift}-\textit{NuSTAR} spectral analysis}\label{spectral_indi}
If, as discussed in the introduction, the inner parts of the accretion disk change over time, we can also investigate any changes within our \textit{NuSTAR} campaign. We do so by exploring the four \textit{Swift}-\textit{NuSTAR} observations (NU1, NU2, NU3, NU4). Our analysis is based on the previous best fit model with a cut-off power law. We keep the absorption parameters and the high-energy cut-off fixed at their best fit values from the fit to the average spectra as we expect them to be unchanged and they are hard to constrain from the individual spectra. The black hole spin $a$, inclination $i$, and iron abundance are linked between the different observations but allowed to vary. For the distant reflection, we assume it remains the same for all observations and use the average photon index for the \texttt{pexmon} model. This results in a very good fit $\chi^2=1836.4/1826$. The results for all spectral parameters are shown in Table~\ref{obs_res}. The spectral parameters are very similar to what we have found from the average spectrum. However, we are now able to constrain the spin to be $>0.5$. This suggests that there are subtle changes to the reflection spectrum that have been smeared out by averaging the spectra. Despite the likely existence of changes, the inner disk radius and the height of the corona above the accretion disk remain unchanged within the errors. NU2, the brightest observation, shows some indication that the inner radius might be larger than for the other observations, but this is not statistically significant. The best fitting inner radii are $<9$, 15$_{-13}^{+38}$, $<39$, $<25\,R_\text{ISCO}$ or $38, 3-224, 165, 106\,R_\text{g}$ accounting for the spin only being constrained to $a>0.5$.

Therefore, as a final effort to uncover any evolution in the inner radius, we fix the spin $a$ at 0.998 and refit. This gives $R_{in}=<21,24_{-10}^{+24},<31,14_{-9}^{+12}\,R_\text{g}$. Again confirming that the fit prefers a recessed disk for NU2 and possibly also NU4. However all values are still consistent with a single value.   


\begin{table}[ht]
\caption{Best fit parameters from joint \textit{Swift}-\textit{NuSTAR} observation-resolved fits with a model including a cut-off power law continuum, cold and ionized reflection, warm absorption, instrinisic neutral absorption and Galactic absorption. Some parameters were kept fixed to those determined in Section~\ref{spectral_all}; see text for details.}\label{obs_res}
\begin{tabular}{c|c|c|c|c}
Parameter & NU1 & NU2 & NU3 & NU4 \\
\hline \hline $\Gamma$ & 1.92$_{-0.03}^{+0.06}$ & 1.83$_{-0.03}^{+0.02}$ & 1.82$_{-0.02}^{+0.03}$ & 1.86$_{-0.04}^{+0.03}$\\
$N_\text{pex}$ [$10^{-3}$] & \multicolumn{4}{c}{$<1.1$}\\
\hline $h$ [$R_\text{g}$] & 10$_{-5}^{+58}$ & $>8$ & $>18$ & 25$_{-15}^{+49}$ \\
$A_\text{Fe}$ & \multicolumn{4}{c}{$0.9_{-0.1}^{+0.3}$} \\
$a$ & \multicolumn{4}{c}{$>0.5$} \\
$i$ [$^\circ$] & \multicolumn{4}{c}{$39^{+10}$} \\
$R_\mathrm{in} [R_\mathrm{ISCO}]$ & $<9$ & 15$_{-13}^{+38}$ & $<39$ & $<25$ \\
$\log(\xi)$ [erg\,cm\,s$^{-1}$] & 1.5$_{-1.1}^{+0.8}$ & 2.7$_{-0.3}^{+0.1}$ & 2.5$\pm 0.2$ & 2.4$\pm 0.3$ \\
$N_\text{rel}$ [$10^{-4}$] & 2.0$_{-0.9}^{+2.9}$ & 2.3$_{-0.1}^{+0.2}$ & 1.9$_{-0.1}^{+0.2}$ & 2.0$_{-0.1}^{+0.6}$ \\
\hline $c_\text{FB}$ & 1.02$\pm 0.02$ & 1.03$_{-0.02}^{+0.01}$ & 1.01$\pm 0.01$ & 1.00$_{-0.01}^{+0.02}$\\
$c_\text{xrt}$ & 0.91$\pm 0.07$ & 0.87$_{-0.04}^{+0.06}$ & 0.92$\pm 0.06$ & 0.89$\pm 0.05$ \\
\hline \hline $\chi^2$/dof & \multicolumn{4}{c}{1836.4/1826} \\
\end{tabular}
\end{table}

\section{Discussion}

The observed spectra are typical for radio galaxies, with a photon index of $\sim$\,1.8 and very little, if any, cold reflection. Further, we detect a high-energy cut-off in the average \textit{NuSTAR} observation. The value of 183$_{-35}^{+51}$\,keV is in very good agreement with the \textit{BeppoSAX} value of $145^{+294}_{-66}$\,keV \citep{Grandi2005} and the tighter, joint \textit{INTEGRAL}-ISGRI/\textit{Swift}-BAT constraint of $190^{+18}_{-66}$\,keV \citep{Malizia2014a}. This suggests that the high-energy cut-off energy in 4C\,74.26 is not strongly variable. The cut-off value fits well with the measurements of the cut-offs of other radio-loud AGN with \textit{NuSTAR}. Specifically, 3C\,382 was found to have a cut-off of 214$_{-63}^{+147}$\,keV \citep{Ballantyne2014} and 3C\,390.3 was found to have a cut-off energy of 117$_{-14}^{+18}$\,keV \citep{Lohfink2015}.  

Modelling the continuum of 4C\,74.26 directly with a thermal Comptonization model, we are able to constrain the actual electron temperature instead of the high energy cut-off. The value of the electron temperature $kT_e=46_{-11}^{+25}$\,keV is in good agreement with what would be inferred from the high-energy cut-off. Together with the photon index we can also estimate the optical depth via the Compton-$y$ parameter. The resulting estimate for the optical depth is about three. A similarly rather large value was found for 3C\,390.3 \citep[$\tau=4-5$;][]{Lohfink2015}, again reaffirming that the spectra of the two sources are similar.  

We also detect cold excess absorption with a column of $<2\times 10^{21}$\,cm$^{-2}$, which is of similar magnitude to what was found in \citet{Ballantyne2005}. Additionally, a mildly ionized warm absorber improves the fit further. The existence of this warm absorber has been observed previously by \citet{Ballantyne2005} and \citet{Gofford2013}. Both the cold absorber as well as the warm absorber appear to be constant with time, implying little change to the surrounding of the black hole over the past decade.

The high quality spectra allow us to observe ionized reflection in the source (Fig.~\ref{residuals_nustaronly}), 4C\,74.26 is one of the most luminous objects to show such clear relativistic reflection features. Still, constraining the accretion disk parameters is difficult given the weak strength of reflection compared to the continuum ($R_\mathrm{ion}\sim0.5$). The low reflection fraction suggests that the corona could be outflowing and the coronal emission being beamed away from the disk \citep{Beloborodov1999,King2016}. From the reflection spectrum we determine that a high black hole spin is likely for the source, similar to what has been seen in other radio-loud AGN \citep{Lohfink2013,Lohfink2015}. A previous study of the reflection spectrum by \citet{Larsson2008} using the models \texttt{reflion} and \texttt{kdblur} constrained the inclination of the viewing angle of the inner disk to $21_{-6}^{+16}$ degrees and the inner radius to $>25\,R_\mathrm{g}$. Their values are in good agreement with ours if one considers that \texttt{kdblur} assumes maximum spin. In our observations a mildly recessed disk is detected from the average spectrum but no significant evolution of the inner radius is seen. However, tentatively, the disk is more recessed at higher flux, contradicting what would be expected from the jet cycle. This would suggest that another process is causing changes to the central regions, at least on the observed timescales. Without any joint radio monitoring observations, however, it is hard to draw any definite conclusions whether a jet cycle is operating during our observations of 4C\,74.26. It is also plausible that the observed mild disk truncation is actually the result of the aforementioned outflowing corona, which could beam the emission away from the innermost parts of the accretion disk.

Finally we note that there are no signs of a jet in the X-ray spectrum. This is confirmed by the lack of an X-ray excess when comparing to the ratio of radio to X-ray luminosity for radio-quiet sources \citep{Miller2011}.

\acknowledgments
\section*{Acknowledgments}
The authors thank the referee for their useful comments that have helped to improve the manuscript. AL acknowledges support from the ERC Advanced Grant FEEDBACK 340442 and useful discussions with Niel Brandt. JAG acknowledges partial funding provided by the Alexander von Humboldt Foundation. This work made use of data supplied by the UK Swift Science Data Centre at the University of Leicester. This work has made use of data from the NuSTAR mission, a project led by the California Institute of Technology, managed by the Jet Propulsion Laboratory, and funded by the National Aeronautics and Space Administration. This research has made use of the NuSTAR Data Analysis Software (NuSTARDAS) jointly developed by the ASI Science Data Center (ASDC, Italy) and the California Institute of Technology (USA).

\bibliographystyle{apj}

\end{document}